\documentclass[iop,apj,times]{emulateapj}

\usepackage{apjfonts} 

\usepackage{times}
\usepackage{graphicx}
\usepackage{latexsym,amssymb}
\usepackage{amsfonts,exscale}

\DeclareGraphicsExtensions{.pdf,.png,.jpg}

\shorttitle{Quasars deflecting the CMB}
\shortauthors{J.~E.~Geach et al.}

\begin{document}

\title{The halo mass of optically-luminous quasars at $z\approx1$--$2$\\measured via gravitational deflection of the Cosmic Microwave Background}

\author{J.~E.~Geach$^1$, J.~A.~Peacock$^2$, A.~D.~Myers$^3$, R.~C.~Hickox$^4$, M.~C.~Burchard$^1$ \& M.~L.~Jones$^5$\vspace{9pt}}

%could you pls add your affils...
\affil{$^1$Centre for Astrophysics Research, School of Physics, Astronomy \& Mathematics, University of
Hertfordshire, Hatfield, AL10 9AB, UK. j.geach@herts.ac.uk}
\affil{$^2$Institute for Astronomy, University of Edinburgh, Royal Observatory, Edinburgh, EH9 3HJ, UK}
\affil{$^3$Department of Physics \& Astronomy, University of Wyoming, 1000 E. University, Dept. 3905, Laramie, Wyoming 82071}
\affil{$^4$Department of Physics \& Astronomy, Dartmouth College, Hanover, New Hampshire 03755}
\affil{$^5$Center for Astrophysics | Harvard \& Smithsonian, 60 Garden Street, Cambridge, Massachusetts 02138}

\begin{abstract}
    We measure the average deflection of cosmic microwave background photons by quasars at $\langle z \rangle =1.7$. Our sample is selected from the Sloan Digital Sky Survey to cover the redshift range $0.9\leq z\leq2.2$ with absolute {\it i}-band magnitudes of $M_i\leq-24$ ({\it K}-corrected to $z=2$). A stack of nearly 200,000 targets reveals an 8$\sigma$ detection of {\it Planck}'s estimate of the lensing convergence towards the quasars. We fit the signal with a model comprising a Navarro--Frenk--White density profile and a 2--halo term accounting for correlated large scale structure, which dominates the observed signal. The best-fitting model is described by an average halo mass $\log_{10}(M_{\rm h}/h^{-1}M_\odot)=12.6\pm0.2$ and linear bias $b=2.7\pm0.3$ at $\smash{\langle z \rangle =1.7}$, in excellent agreement with clustering studies. We also report of a hint, at a 90\% confidence level, of a correlation between the convergence amplitude and luminosity, indicating that quasars brighter than $M_i\lesssim -26$ reside in halos of typical mass $\smash{M_{\rm h}\approx 10^{13}\,h^{-1}M_\odot}$, scaling roughly as $\smash{M_{\rm h}\propto L_{\rm opt}^{3/4}}$ at $\smash{M_i\lesssim-24}$\,mag,  in good agreement with physically--motivated quasar demography models. Although we acknowledge this luminosity dependence is a marginal result, the observed $M_{\rm h}$--$L_{\rm opt}$ relationship could be interpreted as a reflection of the cutoff in the distribution of black hole accretion rates towards high Eddington ratios: the weak trend of $M_{\rm h}$ with $L_{\rm opt}$ observed at low luminosity becomes stronger for the most powerful quasars, which tend to be accreting close to the Eddington limit.

\end{abstract}

\section{Introduction}

Weak lensing of the cosmic microwave background (CMB) is offering an exciting new probe of galaxies, since the surface of last scattering is a $z\approx1100$ backlight to all structure in the observable Universe, and has been gravitationally deflected (Blanchard \& Schneider\ 1987). Fortuitously, the bulk of the matter responsible for deflecting CMB photons coincides with the peak epoch of galaxy growth at $z\approx2$ (Cooray \& Hu\ 2000). 

Galaxies form at peaks in the matter density field (Press \& Schechter\ 1974; Bardeen et al.\ 1986; Peacock \& Heavens\ 1990), with a formation history that is sensitive to both local and large-scale environment; they are therefore `biased' tracers of the distribution of mass in the Universe (e.g.,\ Baugh et al.\ 1999). Measuring this bias has been the standard means of placing a galaxy population in its cosmological context, since one can estimate the typical dark matter halo mass hosting members of that population through the halo--bias relation (e.g.,\ Sheth \& Tormen\ 1999; Tinker et al.\ 2010). The two-point correlation function has been the tool of choice for estimating galaxy bias for decades (e.g., Peebles\ 1980), but the amplitude of the cross-power signal measured between maps of the projected density of galaxies and maps of the convergence of the CMB lensing field also offers a clean measurement of the bias (e.g.,\ Bleem et al.\ 2012; Sherwin et al.\ 2012; Geach et al.\ 2013; DiPompeo et al.\ 2017). Inferring the bias by cross-correlation with the CMB lensing field has an advantage over the bias inferred from quasar--galaxy cross-correlation, in that there is no contribution from any stochastic component of quasar or galaxy bias (Dekel \& Lahav\ 1999). 

Cross-correlation is often performed in Fourier space, and in real space it is possible to detect the average deflection of CMB photons at the positions of galaxies in `stacks' of the convergence field around single halos\footnote{Stacking is essentially just the real space equivalent of cross-correlation, but in theory, sensitive and high-resolution CMB observations should be able to detect the lensing signal due to individual deflectors, a feat that could potentially be achievable with the Atacama Large Millimeter/submillimeter Array (Holder \& Kosowsky 2004).} (Baxter et al.\ 2015; Madhavacheril et al.\ 2015; Geach \& Peacock\ 2017). An important distinction of this approach compared to {\it galaxy} weak lensing studies (e.g.,\ Massey et al.\ 2007) is that it can be applied at high redshift ($z>1$), where the background galaxies at even greater distances are too faint and poorly resolved to permit useful measurements of their lensing distortions.

Quasars are visible over the bulk of cosmic history, making them excellent tracers of large scale structure, and represent an important phase in galaxy evolution (e.g.,\ Silk \& Rees\ 1998). As such, analysis of the clustering of quasars has a rich history (Rees \& Sciama\ 1967; Shanks et al.\ 1983; Croom et al.\ 2005; Eftekharzadeh et al.\ 2015; Laurent et al.\ 2017). Quasar surveys have played a vital role through ground-breaking observational campaigns such as the 2dF QSO Redshift Survey (2QZ, Boyle et al.\ 2000), and continue to play a role with the Sloan Digital Sky Survey (SDSS)-III Baryon Oscillation Spectroscopic Survey (BOSS, Dawson et al.\ 2013) and SDSS-IV extended--BOSS (eBOSS, Dawson et al.\ 2016).

Here we measure the halo mass of quasars through a CMB lensing stacking analysis of optically selected quasars at $0.9\leq z \leq 2.2$. Unlike previous works that have studied quasar--CMB lensing cross-correlation (e.g.,\ Sherwin et al.\ 2012; Geach et al.\ 2013; DiPompeo et al.\ 2017), here we use a spectroscopic sample of optically-selected quasars with a carefully designed selection function, for which the redshift distribution is exactly known. Moreover, we present an analysis of the luminosity dependence of the quasar halo mass. In Section 2 we describe our methods, and present the results and analysis in Section 3. We discuss the results and present our conclusions in Sections 4 and 5. Throughout we assume a `{\it Planck} 2015' cosmology with $\Omega_{\rm m}=0.307$, $\Omega_{\rm \Lambda}=0.693$, $h=H_0/100\,{\rm km\,s^{-1}\,Mpc^{-1}}=0.677$, $\sigma_8=0.8159$, $n_{\rm s}=0.9667$ (Planck Collaboration\ 2015). Unless otherwise stated, all distances are proper lengths, not comoving.

\section{Methods}

\subsection{Lensing map}

We use the 2018 release of the {\it Planck} lensing convergence baseline map including CMB-only estimates of the lensing signal to scales of $\ell=4096$\footnote{Lensing data products used in this work are available from the {\it Planck} Legacy Archive: http://pla.esac.esa.int/pla} (Planck Collaboration\ 2018), adopting the minimum variance (MV) lensing estimator (Carron \& Lewis\ 2017). We impose an additional filtering step by zeroing $a_{\ell m}$ amplitudes for $\ell<100$ (J. Carron\,private communication). 

We do not use the joint CMB--CIB lensing estimate released by {\it Planck} because the joint lensing reconstruction will gain a direct contribution from thermal dust emission due to the quasars themselves, or due to galaxies in their local environment (e.g.,\ Stevens et al.\ 2010), and so will not give a pure indication of gravitational deflection at the location of the quasars. We refer the reader to the {\it Planck} 2018 lensing paper {\sc viii} for a thorough description of the construction of the {\it Planck} lensing products.

\subsection{Quasar sample}

We use the 14th Data Release (DR14) of the SDSS quasar catalog (P\^aris et al.\ 2018), which includes sources from the eBOSS (Dawson et al.\ 2016) component of SDSS-IV, as well as pre-BOSS (Data Release 7) targets for a total of 526,356 sources. The original selection of eBOSS `{\sc core}' quasars (Myers et al.\ 2015) was designed to return a statistically uniform sample over $0.9\leq z\leq2.2$ for the purpose of detecting the signature of baryonic acoustic oscillations in the correlation function. From the DR14 catalog we select quasars in the range $0.9\leq z \leq 2.2$ and apply a luminosity cut of $M_i(z=2)\leq-24$, where the {\it i}-band absolute magnitude is {\it K}-corrected to $z=2$ following Richards et al.\ (2006). The optical luminosity cut ensures roughly uniform completeness across $0.9\leq  z \leq 2.2$. Our only other selection criterion is to reject quasars that are detected at 1.4\,GHz by the Faint Images of the Radio Sky at Twenty-Centimeters (FIRST) survey (Becker et al.\ 1995). It is necessary to remove radio-loud quasars from the sample, since this radio-selected population may have rather different clustering properties to optically-selected quasars. Note that, while our redshift selection is designed to match the {\sc core} target range, our sample also includes pre-BOSS targets in the same redshift and luminosity cut and so is not strictly homogeneous. Figure\ 1 shows the redshift and luminosity--redshift plane distribution of the quasar sample, and a comparison to the lensing kernel $W^\kappa$ (e.g.,\ Cooray \& Hu\ 2000). The total sample size for the following stacking analysis is 197,784 quasars.

\begin{figure}
    \centering
    \includegraphics[width=0.45\textwidth]{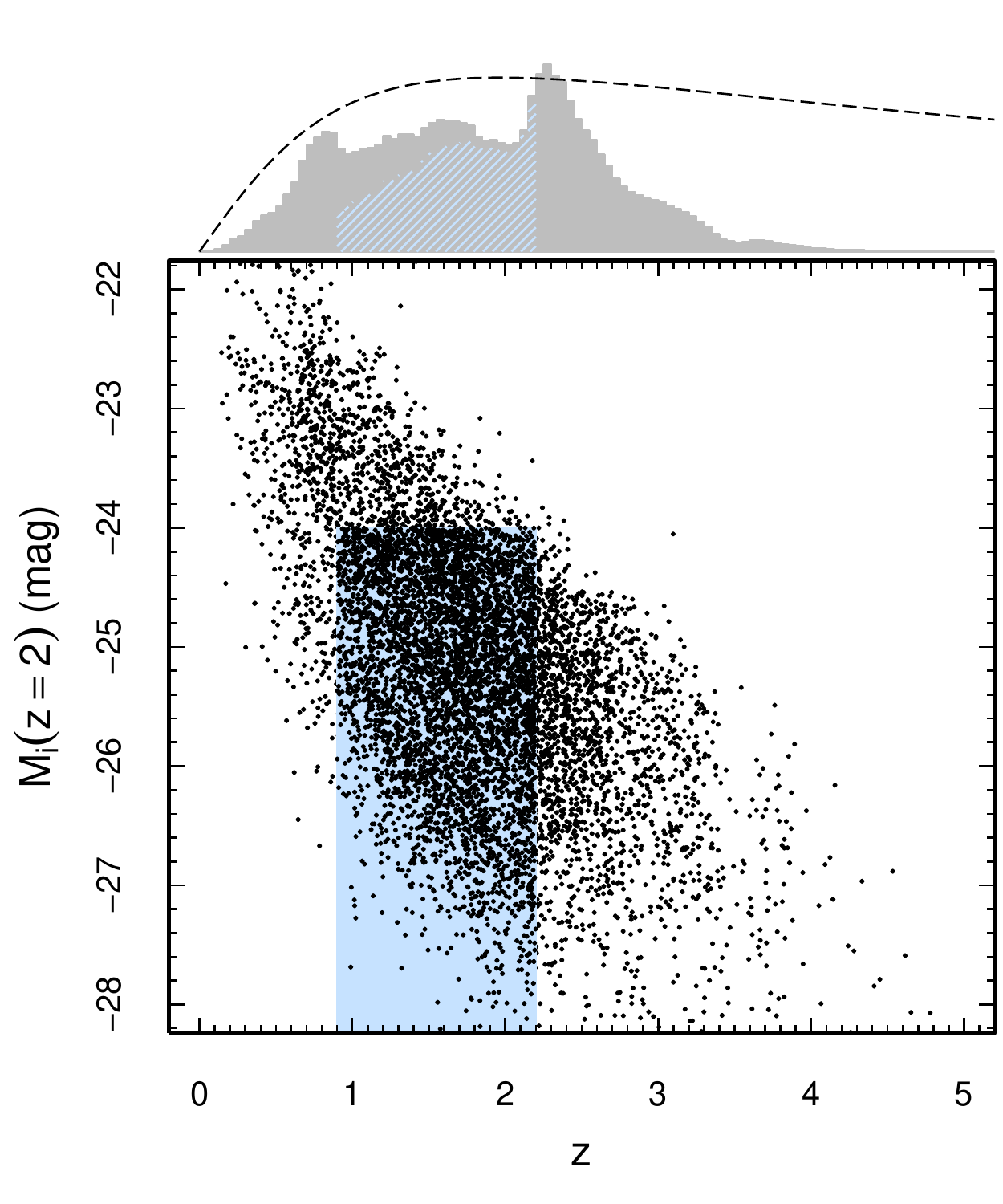}
    \caption{Selection of quasars. The blue shaded rectangle shows our selection of $M_i(z=2)\leq-24$, $0.9\leq z\leq 2.2$ quasars from the SDSS DR14 quasar catalog, designed to be an approximate match to a luminosity-complete sample of `{\sc core}' quasars. The histogram shows the redshift distribution of the full catalog and our selection, which has a median of $\langle z \rangle =1.7$. The total sample contains 197,784 quasars. The dashed line shows the shape of the CMB lensing kernel, plotted as the conventional $({\rm d}\chi/{\rm d}z)W^\kappa$.}
\end{figure}

\begin{figure*}
    \centering
    \centerline{\includegraphics[width=\textwidth]{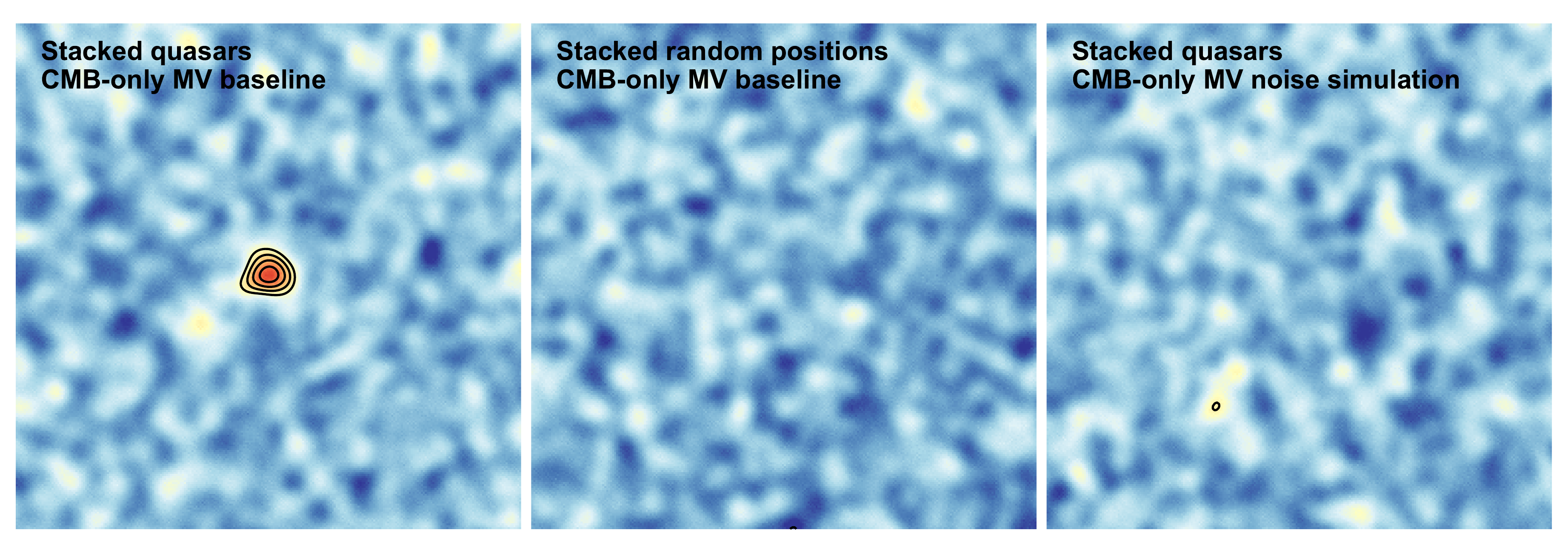}}
    \caption{Stacks spanning 5\,degrees at the positions of quasars. (left) the stack of the baseline MV convergence map, (center) an identical stack after randomly shifting quasar positions and (right) an identical stack in one of the noise-only MV simulations. The contours are at levels of 3, 4, 5... $\sigma$. The peak convergence is $\smash{\langle \kappa \rangle = (2.4\pm0.3)\times10^{-3}}$.}
\end{figure*}

\subsection{Stacking}

Following the method presented in Geach \& Peacock\ (2017), we project the convergence map onto a tangential (Zenithal Equal Area) flat sky projection spanning 5\,degrees at the position of every quasar, preserving any masking present in the convergence map. We adopt a grid size of $256\times256$ pixels and use nearest neighbour interpolation in the re-projection. The stacked map is simply the pixel-wise mean convergence where masked pixels are ignored. The input convergence map itself is generated from the set of $100\leq \ell \leq 4096$ $a_{\rm \ell m}$ coefficients (using the {\sc healpy}\footnote{https://github.com/healpy/healpy} {\it alm2map} function), where the mean field is removed. Since high-$\ell$ modes are noisy, to improve signal-to-noise at the angular scales of interest, we smooth the map with a Gaussian kernel with  a {\sc fwhm} of 15$'$. This stack is repeated for the same quasar sample in each of 300 noise realizations of the convergence map, where each noise realization is filtered in an identical manner to the data. We use the variance of this ensemble of stacks to evaluate the significance of the signal. As an additional check, we also perform stacks in the real map after randomly shifting the right ascension of each quasar by 2 to 15\,degrees. 

\section{Results and analysis}
The stack is presented in Figure\ 2, showing a significant detection of the average convergence with a peak $\langle \kappa \rangle = (2.4\pm0.3)\times10^{-3}$, and confirming a null detection in both the random position stack and noise-only stack. We detect no residual in the equivalent stack of the {\sc smica} temperature map at the positions of the same targets. For {\it Planck} data, Geach \& Peacock\ (2017) showed that, even in the case of clusters of galaxies, where a strong thermal Sunyaev-Zel'dovich (tSZ) signal remains in the temperature map, such a residual distortion has a negligible effect on the lensing estimate. Therefore, we do not consider any low-level tSZ signal associated with quasars to be biasing the lensing estimate.

\subsection{Average halo mass}

To evaluate the average halo mass from the stacked convergence signal we adopt the same procedure as Geach \& Peacock (2017), which is summarized here. First, we assume a Navarro-Frenk-White (NFW) density profile (Navarro, Frenk \& White\ 1997) for halos hosting quasars 
\begin{equation}
\rho(x) = \frac{\rho_s}{x(1+x)^{2}},
\end{equation} 
where $x=r/r_s$ and $r_s$ is a scale radius. We adopt $r_{\rm s}=r_{200}/c$,
where $r_{200}$ is the radius at which the enclosed mass density is
equal to 200 times the {\it mean} density of the Universe and $c$ is the halo concentration parameter. For the latter we assume the scaling of Dutton \& Macci\`o\ (2014). The projected density is then
\begin{equation}
\Sigma(R) = 2 \int^{\infty}_R \frac{r\,\rho(r)}{\sqrt{(r^2-R^2)}}\;{\rm d}r.
\end{equation}
The lensing convergence for the 1-halo term is defined as the ratio of projected mass surface density to the critical surface density 
\begin{equation}
\kappa_1(R) = \frac{\Sigma(R)}{\Sigma_{\rm crit}},
\end{equation}
where the critical surface density is
\begin{equation}
\Sigma_{\rm crit} = \frac{c^2}{4\pi G}\frac{D_{\rm OS}}{D_{\rm OL}D_{\rm LS}}. 
\end{equation}
Here $D$ is the angular diameter distance, and the subscripts {\sc os}, {\sc ol} and {\sc ls} denote the distance between observer--source, observer--lens and lens--source, where the source is the surface of last scattering at $z=1100$. Note that we assume quasars are located at the centers of halos, and therefore apply no off-centering smoothing to the predicted convergence profile. 

Equations 1--4 allow us to model the convergence profile for individual halos, but we also consider a 2-halo term to describe a contribution to the lensing signal due to correlated large scale structure, using Limber's approximation (e.g.,\ Oguri \& Hamana\ 2011):
\begin{equation}
\kappa_{\rm 2}(\theta) = \frac{\bar{\rho}(z)}{(1+z)^3\Sigma_{\rm
    crit}D^2(z)}\int
\frac{\ell\, {\rm d}\ell}{2\pi}J_0(\ell\theta)\,b_{\rm h}\,\Delta(k,z),
\end{equation}
where $J_0$ is the zeroth order Bessel function, $D(z)$ is the
angular diameter distance, $\Delta(k,z)$ is the linear matter power spectrum
(with comoving wavenumber $k=\ell/(1+z)D$), following Eisenstein \& Hu\ (1999), $\bar{\rho}(z)$ is the average density of
the Universe at $z$ and $b_{\rm h}$ is the linear bias for a halo of mass
$M_{\rm h}$ (we take the bias to be constant across scale and redshift). We assume $b_{\rm h}=f(\nu)$, where $\nu$ is the ratio of the critical
threshold for spherical collapse to the root mean squared density
fluctuation for a halo of mass $M_{\rm h}$, $\nu=\delta_c/\sigma(M,z)$, and $f$ is the functional fit of Tinker et al.\ (2010). Finally, the model stacked convergence for a sample with normalized redshift distribution ${\rm d}n/{\rm d}z$ is 
 \begin{equation}
     \langle\kappa\rangle = \int {\rm d}z~(\kappa_1+\kappa_2)~{\rm d}n/{\rm d}z.
 \end{equation}

To find the best-fitting mass $M_{\rm h}$ we use Equations 1--5 to generate a set of lensing profiles for a wide range of halo mass on the same grid as the stacked map, applying the same filtering as the data, and then fit for the peak convergence amplitude with $M_{\rm h}$ as a free parameter. To estimate the 1$\sigma$ uncertainty on $M_{\rm h}$, we repeat the fit 300 times, each time adding a noise map to the data. The noise maps are derived from the same stacking procedure that was used on the real data, but replacing the true convergence map with the set of signal-free noise realisations released as part of the {\it Planck} lensing package. We take the standard deviation of the ensemble of best fits as the 1$\sigma$ uncertainty on the halo mass, finding $\log_{10}(M_{\rm h}/h^{-1}M_\odot)=12.6\pm 0.2$. Figure\ 3 shows the radial profile of the stacked lensing signal and best-fitting model.

It is impressive, and perhaps surprising, that the stacked lensing signal is able to detect this small-scale structure in the convergence field, given that the focus of the CMB reconstruction was on the large-scale convergence field. However, Geach \& Peacock (2017) verified by direct simulation that the algorithm being used allows the convergence field to be recovered in an unbiased manner even in the cores of halos in the {\it Planck} maps.

\begin{figure}
    \centering
    \centerline{\includegraphics[width=0.5\textwidth]{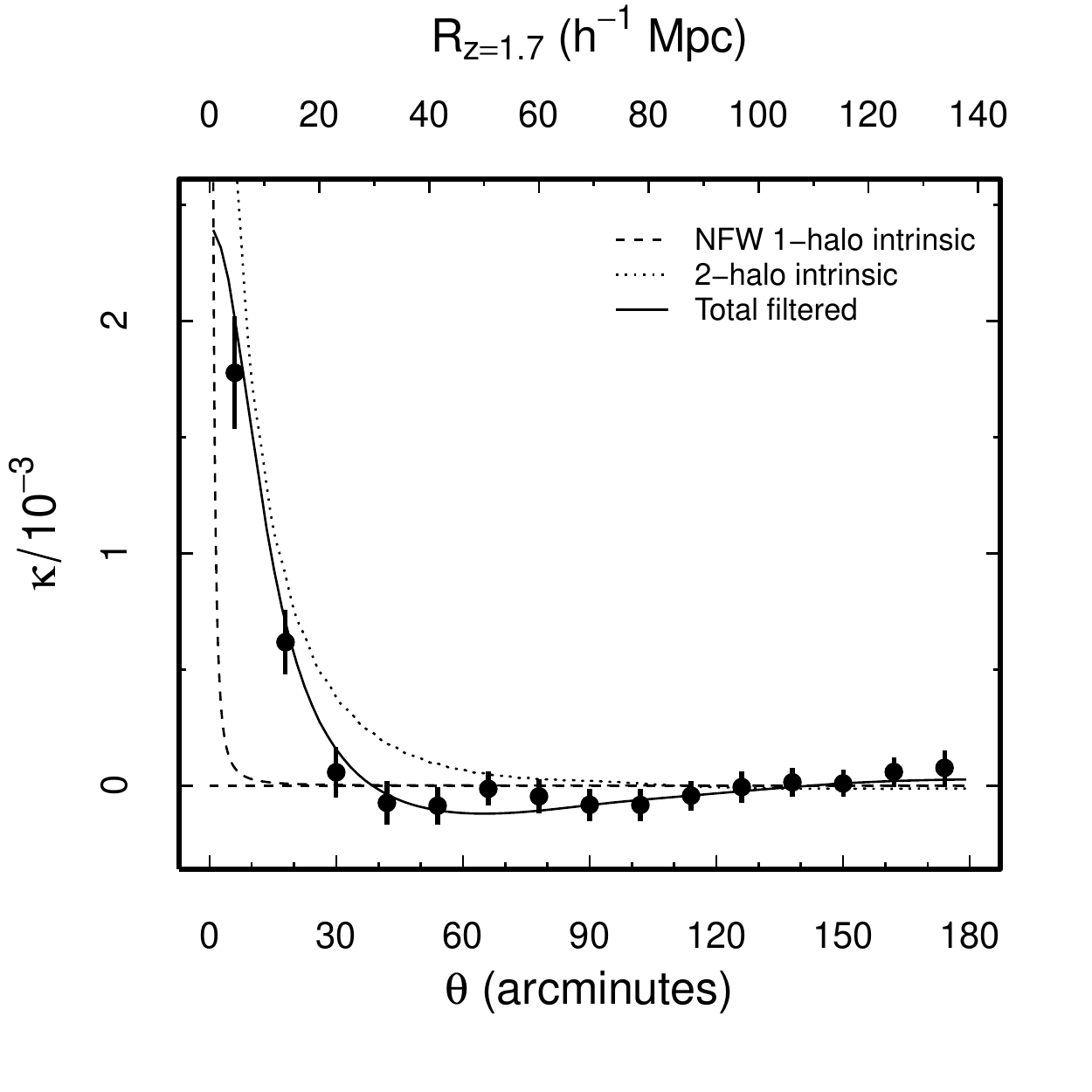}}
    \caption{The radial profile of the quasar stacked convergence. Error bars represent the 1$\sigma$ scatter in each bin across 300 identical stacks in the noise maps. The solid line shows the best fitting model. The model includes the 1- and 2-halo contribution to the lensing signal (Section 3.1) and has been filtered in an identical manner to the data. The dot and dot-dash lines show the intrinsic (unfiltered) lensing profiles for the 1- and 2-halo components. The top abscissa shows the physical scale at the median redshift $\langle z \rangle =1.7$.}
\end{figure}

\subsection{Luminosity dependence}

We split the sample into three bins of luminosity: $-24\leq M_i<-25$ (85,450 quasars), $-25\leq M_i<-26$ (67,079 quasars) and $-26\leq M_i<-27$ (35,377 quasars), and stack as above. We measure significant peak convergence amplitudes of $\langle \kappa \rangle =(2.4\pm0.5)\times10^{-3}$, $\langle \kappa \rangle =(2.5\pm0.6)\times10^{-3}$ and $\langle \kappa \rangle =(3.7\pm0.6)\times10^{-3}$ corresponding to $\log_{10}(M_{\rm h}/h^{-1}M_\odot)=12.5\pm0.3$, $12.6\pm0.3$ and $13.0\pm0.2$ respectively. Note we have used the exact redshift distributions for each cut when calculating the halo model; the median redshifts are $\langle z\rangle=1.61$, $\langle z\rangle=1.68$ and $\langle z\rangle=1.75$. To estimate the significance of the trend between the lensing convergence and optical luminosity, we fit a simple linear model $\kappa=aM_i + b$, finding $a=-0.62\pm0.39$, $b=-12.94\pm9.86$, which is preferred at a 1.6$\sigma$ (90\% confidence) level compared to the null hypothesis of no trend. We take this as tentative evidence for a luminosity dependence of the quasar halo mass. The predicted peak convergence amplitude for a halo of mass $\log_{10}(M_{\rm h}/h^{-1}M_\odot)=12.5$ varies by just 3\% for the different ${\rm d}n/{\rm d}z$ in each luminosity bin, and the expected evolution in the quasar bias is mild over this range (Croom et al.\ 2005; Laurent et al.\ 2017). 

In Figure~4 we plot the halo mass as a function of optical luminosity (where $\log_{\rm 10}(L_{\rm opt}/{\rm W})=-0.4(M_i-72.5)$, e.g.\ Shen et al.\ 2009). These measurements hint at an upturn in the average halo mass for the most luminous quasars, with host masses of order $\smash{M_{\rm h}\approx 10^{13}h^{-1}M_\odot}$ for quasars with  $M_i\approx -26$, approximately a factor of 3 higher than those with $M_i\approx -24$. However, we again caution this remains a tentative result.

\begin{figure}
    \centering
    \centerline{\includegraphics[width=0.5\textwidth]{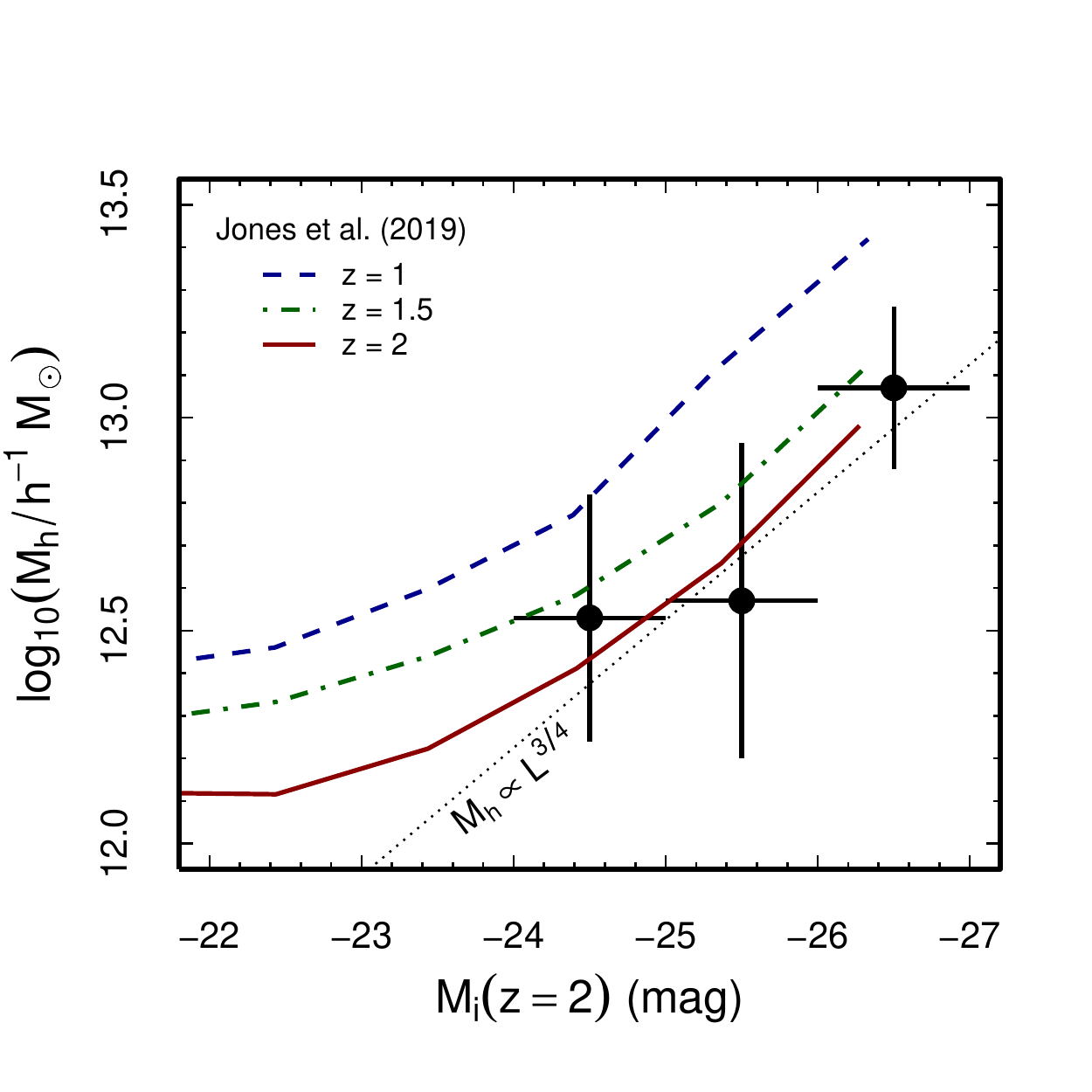}}
    \caption{Average halo mass as a function of optical luminosity based on stacks in luminosity bins. The dotted line shows a scaling between halo mass and optical luminosity $\smash{M_{\rm h}\propto L^{3/4}}$ (e.g.,\ Silk \& Rees\ 1998). The other lines show the semi-numerical model of Jones et al.\ (2019, submitted) for redshifts $z=1,1.5$ and $2$, which are in good agreement with the data for luminosities brighter than $M_i<-24$\,mag, with an upturn at high luminosities reflecting the fact that these quasars are powered by black holes accreting close to the Eddington limit, and therefore have a narrower distribution of Eddington ratios, resulting in a stronger $M_{\rm h}$--$L_{\rm opt}$ correlation at the highest luminosities.}
\end{figure}

\section{Discussion}

It is important to note what signal has actually been detected. As Figure\ 3 illustrates, on scales of $\theta>15'$ we are sensitive to the 2-halo contribution to the lensing signal, with the bulk of the 1-halo signal filtered out of the {\it Planck} lensing estimate. The amplitude of the 2-halo term depends on the halo bias, which links the power spectrum of halos of a given mass to the underlying matter power spectrum (Equation\ 5). With a constraint on the bias, one can estimate the halo mass through the halo--bias function, where we have have used the commonly used calibration of Tinker et al.\ (2010) based on {\it N}-body simulations (although we note that some caution must be exercised in applying this calibration for small halo masses due to the potential influence of baryonic effects that are not captured in the simulation). It is clear that, rather than directly constraining the average halo mass due to the lensing signal of the host halo, we have detected the signature of lensing by matter correlated with quasar halos on scales of 10\,Mpc. So ultimately, we have measured the quasar halo bias, which is $b_{\rm h}=2.7\pm 0.3$ at $\langle z\rangle=1.7$ for quasars with $M_i<-24$\,mag.

Previous measurements of the quasar bias have been generally derived from the two-point auto-correlation function, measured in angular projection or in redshift space. Although not without its own particular set of systematic effects (e.g.,\ van Engelen et al.\ 2014), using the CMB lensing field  potentially offers a cleaner approach compared to traditional clustering analyses, in which the systematics (related to incompleteness, random catalog generation, masking, etc.) must be carefully accounted for (e.g.,\ Reid et al.\ 2016). With this in mind, how does the lensing estimate of the quasar bias compare to clustering analyses? 

Croom et al.\ (2005) used the 2QZ survey to establish a model for the evolution of the quasar bias, $b_{\rm h}(z) = (0.53 \pm 0.19) + (0.289 \pm 0.035)(1 + z)^2$. This model predicts a mean $\langle b_{\rm h} \rangle = 2.63\pm0.32$ for $z=1.7$, consistent with our measurement. More recently, Laurent et al.\ (2017) measured the bias of eBOSS quasars through the across the same redshift range as considered here, finding $b_{\rm h} = 2.45 \pm 0.05$ at $z=1.55$ and an evolution that can be modelled as $b_{\rm h}(z) =(0.278\pm0.018)\left[(1+z)^2 - 6.565\right] + (2.393\pm0.042)$. This yields $b_{\rm h}=2.59\pm 0.04$ for $z=1.7$, again consistent with our lensing-derived estimate of the bias at this epoch.

While our measurement of the bias is not as precise as previous estimates, our analysis has revealed mild evidence for a correlation between quasar luminosity and halo mass. Previously, it has proven challenging to establish whether such a dependence exists at all (e.g.\ Myers et al.\ 2007; Shen et al.\ 2009; 2013). In Figure\ 4 we compare our measurement of $M_{\rm h}$ as a function of quasar luminosity to the predictions of the model of black hole growth in galaxies by Jones et al.\ (2019, submitted). This model utilizes a semi-numerical treatment for galaxy formation based on the Millennium dark matter simulation (Mutch et al.\ 2013) and adds a prescription for black hole growth in which all black holes have instantaneous accretion rates drawn from a broad Eddington ratio distribution that spans several orders of magnitude, with a cutoff near the Eddington limit (e.g., Hickox et al.\ 2014; Jones et al.\ 2016). We compute the average predicted $M_{\rm h}$ in bins of optical quasar luminosity at $z=1$, $1.5$ and $2$, and obtain good agreement with our CMB lensing results (Figure\ 4), while also reproducing the previously observed weak dependence of $M_{\rm h}$ on luminosity at fainter $M_i$ (e.g.,\ Shen et al.\ 2013).

\section{Conclusion}

What can one conclude? There are three important points to note. First, our study supports the view that quasars at $z\approx1$--$2$ with $M_i<-24$\,mag reside in dark matter halos of characteristic mass $M_{\rm h}\approx 3\times10^{12}\,h^{-1}M_\odot$, arriving at this result using a technique that is different to the traditional two-point quasar-quasar or quasar-galaxy auto- and cross-correlation measurements (and we note that the excellent consistency between the approaches is encouraging). The second point is that we have revealed tentative evidence that the most luminous quasars reside, on average, in more massive halos, with a mass dependence that scales roughly as $\smash{L_{\rm opt}^{3/4}}$ (e.g.\ Silk \& Rees\ 1998) for quasars brighter than $M_i<-24$\,mag, and below which the $M_{\rm h}$--$L_{\rm opt}$ scaling is predicted to flatten. The upturn in the $M_{\rm halo}$ at high luminosity occurs because these most luminous quasars are necessarily accreting near the Eddington limit and so have a narrow range of Eddington ratios. This leads to a strong correlation between luminosity and black hole (and thus dark matter halo) mass, while lower luminosity quasars span a broader range of Eddington ratios and thus exhibit weaker correlations between luminosity and mass. The final point is that, unlike clustering studies that only ever trace the underlying matter field, here we have {\it directly} measured the total projected mass density around quasars on scales of order \smash{10\,$h^{-1}$\,Mpc}. The fact that this signal can be accurately modelled with the simple linear bias model should be seen as a success in our understanding of how galaxies trace the large-scale distribution of matter in the Universe.

\section*{Acknowledgements}

We thank the two anonymous referees whose comments have improved this work. JEG is supported by the Royal Society through a University Research Fellowship. JAP is supported by the European Research Council under grant number 670193 (the COSFORM project). ADM acknowledges support by the National Science Foundation (NSF) through grant number 1616168 and by the U.S. Department of Energy, Office of Science, Office of High Energy Physics, under Award Number DE-SC0019022. RCH acknowledges support from the NSF through grant numbers 1515364 and 1554584, and from NASA through grant number NNX16AN48G. MLJ acknowledges support from NASA-MUREP under grant number NNX15AU32H. The authors wish to thank Julien Carron, Antony Lewis and Gil Holder for invaluable discussions. This work made use of the COsmology, haLO, and large-Scale StrUcture toolS ({\sc colossus}) Python package (Diemer\ 2017). Some of the results in this paper have been derived using the HEALPix package (G{\'o}rski et al.\ 2005).

\end{document}